\documentclass[aps,pra,twocolumn,showpacs,superscriptaddress,longbibliography]{revtex4-1}
\usepackage{graphicx}
\usepackage{adjustbox}
\usepackage{amsmath}
 \usepackage{booktabs}
 \usepackage{multirow}
\usepackage{amssymb}

\usepackage{epstopdf}

\usepackage{amsfonts}

\usepackage{dcolumn}

\usepackage{multirow}
\usepackage[utf8]{inputenc}
\usepackage{esint}
\usepackage{float}
\usepackage{breakurl}
\usepackage{hyperref}
\hypersetup{colorlinks,citecolor=blue}

\makeatletter
\@ifundefined{textcolor}{}
{
 \definecolor{BLACK}{gray}{0}
 \definecolor{WHITE}{gray}{1}
 \definecolor{RED}{rgb}{1,0,0}
 \definecolor{GREEN}{rgb}{0,1,0}
 \definecolor{BLUE}{rgb}{0,0,1}
 \definecolor{CYAN}{cmyk}{1,0,0,0}
 \definecolor{MAGENTA}{cmyk}{0,1,0,0}
 \definecolor{YELLOW}{cmyk}{0,0,1,0}
}

\usepackage{subfigure}\usepackage{epsfig}\usepackage{amsfonts}\usepackage{mathrsfs}
\usepackage[toc,page,title,titletoc,header]{appendix}
\usepackage{amsmath}

\makeatother

\begin{document}
\title{Many-body localization of one-dimensional degenerate Fermi gases with cavity-assisted non-local quasiperiodic interactions}
\author{Jianwen Jie}
\email{Jianwen.Jie1990@gmail.com}
\affiliation{Shenzhen Institute for Quantum Science and Engineering, Southern University of Science and Technology, Shenzhen 518055, China}
\affiliation{International Quantum Academy, Shenzhen 518048, China}
\affiliation{Guangdong Provincial Key Laboratory of Quantum Science and Engineering, Southern University of Science and Technology, Shenzhen 518055, China}
\author{Qingze Guan}
\email{gqz0001@gmail.com}
\affiliation{Homer L. Dodge Department of Physics and Astronomy, The University of Oklahoma,
440 West Brooks Street, Norman, Oklahoma 73019, USA}
\affiliation{Center for Quantum Research and Technology, The University of Oklahoma, Norman, Oklahoma 73019, USA}
\affiliation{Department of Physics and Astronomy, Washington State University, Pullman, Washington 99164-2814, USA}
\author{Jian-Song Pan}
\email{panjsong@scu.edu.cn}
\affiliation{College of Physics, Sichuan University, Chengdu 610065, China}
\affiliation{Key Laboratory of High Energy Density Physics and Technology of Ministry of Education, Sichuan University, Chengdu 610065, China}

\date{today}

\begin{abstract}
The localization properties of one-dimensional degenerate Fermi gases with cavity-assisted non-local quasiperiodic interactions are numerically studied. Although the cavity-induced interaction is typically nonlocal, it is proved that the {\color{black} eigenstate thermalization hypothesis (ETH)} is still applicable in our system depending on the system parameters. We also find the segment of the spectrum corresponding to infinite {\color{black} effective} temperature varies for different system parameters, which indicates the spectral range employed in the spectral statistical analysis should be varied accordingly. The features of many-body localization (MBL) are numerically identified by analyzing the spectral statistics and the entanglement entropy using exact diagonalization. These features are further confirmed by our time evolution results. In addition, the number of cavity photons are found stable over long time dynamics in the MBL phase. Such a feature can not only be utilized to nondestructively diagnose the MBL phase by monitoring the number of leaking photons from the cavity, but leveraged for constructing a device to produce a stable number of photons.
\end{abstract}
\maketitle

\section{Introduction}
The enduring research interests in exploring localized phases are triggered by the seminal work of Anderson~\cite{anderson1958absence} in which the localization of a free particle wave function in a random potential was characterized.  Interactions can easily break the integrability of a quantum system and fuzz the concept of constructive interference, a key mechanism of leading single-particle localizations. Thus, the localization of an interacting many-body system~\cite{basko2006dm, basko2008problems}, known as many-body localization (MBL), is highly nontrivial. Understanding the effects of different types of interactions is a basic task in the study of MBL nowadays.

In contrast to MBL in systems with short-range interactions, the phase boundary of MBL in systems with long-range interactions has been actively under study today. Intuitively, it is expected that the formation of localization will be suppressed with increasing interaction range~\cite{georgeot1998integrability}. For power-law interactions, the critical range
can be estimated based on perturbation analysis in the strong-disorder limit~\cite{burin1994low, esquinazi2013tunneling, burin2006energy, yao2014many}, which has been confirmed numerically~\cite{li2016many}. For completely non-local interactions, as are usually found in cavity-atom-hybrid systems,
it is shown that the MBL phase of Bosons with cavity-assisted quasiperiodic non-local interactions is unstable (stable) in the thermodynamic limit with (without) the condition that the interaction strength scales with the system size~\cite{kubala2021ergodicity}. Compared to intra-cavity ultracold Bose gases~\cite{baumann2010dicke,lin2019superfluid, lin2021mott}, degenerate Fermi gases (or hard-core Bosons~\cite{rylands2020photon}) coupled to cavity photons show distinct atom-photon scattering behaviors. For example, the critical atom-photon coupling strength for the superradiant phase transition vanishes in a cavity-confined one-dimensional Fermi gas given the cavity photon wavelength commensurate with the Fermi wavelength~\cite{piazza2014umklapp,keeling2014fermionic,chen2014superradiance,pan2015topological,mivehvar2017superradiant,yu2018topological,zhang2021observation}.

Inspired by recent experimental realizations of intra-cavity Fermi gases~\cite{roux2020strongly,zhang2021observation,roux2021cavity}, this work studies the localization of one-dimensional Fermi gases with long-range quasiperiodic interactions.
We confirm that the {\color{black}ETH} is still applicable for the intra-cavity Fermi gas, although the system is subject to inter-particle interactions of boundless ranges.
The features of MBL are numerically identified by analyzing the spectral statistics and the entanglement entropy using exact diagonalization.
One observation is that the segment of the spectrum corresponding to infinite temperature moves toward the high energy direction with increasing interaction strength
which implies that the states involved in the spectral statistical analysis should be adjusted accordingly.
Finally,
we find that the number of cavity photons in the MBL phase is almost stationary with time evolution.
Such a feature can not only be utilized to nondestructively diagnose the MBL phase by monitoring the number of leaking photons from the cavity, but leveraged for producing a stable number of cavity photons given a carefully designed atomic initial state.

In the following, we introduce our model in Sec. \ref{model}. The applicability of {\color{black}ETH} and the segment of the spectrum corresponding to infinite temperature are discussed in Sec. \ref{secETH}.
In Sec. \ref{secMBL} we show the exact diagonalization results in identifying the MBL phase. Sec. \ref{timevolution} presents the time evolution results. Finally, a summary is given in Sec. \ref{secVI}.

\begin{figure}[tbp]
\includegraphics[width=8.5cm]{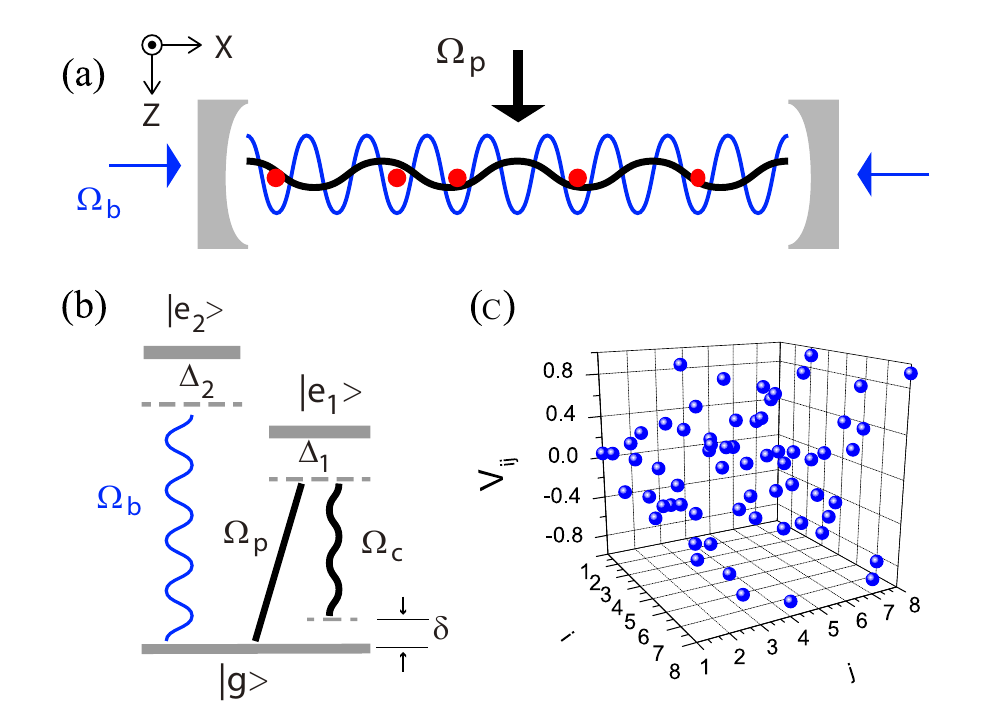}
\caption{(Color online) (a) Illustration of the proposed model: a quasi-one-dimensional Fermi gas is trapped in an optical cavity. The level configuration of the atoms is shown in (b). A laser, {\color{black}$\Omega_b$},  with far-off resonant frequency with respect to the cavity mode {\color{black}$\Omega_c$}, is applied to generate a back-ground lattice. The cavity-atom hybridized system is driven by the nearly resonant transverse laser $\Omega_p$. The wavelengthes of $\Omega_{p}$ and $\Omega_{b}$ are incommensurate which gives rise to the quasiperiodic cavity-assisted interaction (derived in Appendix \ref{appendix1}). A typical distribution of the quasiperiodic interaction $V_{ij}$ is illustrated in (c).}
\label{fig:model}
\end{figure}

\section{Model}\label{model}
Our system consists of a quasi-one-dimensional degenerate Fermi gas with three atomic energy levels ($|g\rangle, |e_1\rangle,$ and $|e_2\rangle$) confined in an optical cavity in the $x$-direction [see Fig.~\ref{fig:model}(a-b)].
A longitudinal standing-wave $\Omega_b\propto\cos(k_0x)$ is present in the cavity to form a background optical lattice (the blue curves in Fig.~\ref{fig:model}).
An external driving laser field $\Omega_p\propto \cos(k z)$  is applied along the $z$-direction to nearly resonantly couple the cavity mode $\Omega_c\propto\cos(kx+\phi)$ via a Raman process between states $|g\rangle$ and $|e_1\rangle$ [see Fig.~\ref{fig:model}(b)].
To obtain randomness in the effective atom-atom interaction term,
the wave number $k_0$ of $\Omega_b$ is adjusted to be incommensurate with that of the cavity mode $\Omega_c$, i.e., irrational value of $k/k_0$.
All the possible one-photon processes involving the optical lattice beam, the cavity mode, and the driving laser beam are far detuned from any of the atomic resonance transitions,
ensuring that the high-lying states $|e_{1,2}\rangle$ can be adiabatically eliminated~\cite{kubala2021ergodicity,zhou2011cavity}.

The Hamiltonian of our model reads (see Appendix \ref{appendix1} for the detailed derivation)
\begin{equation}\label{eq:TIH_without_a}
\hat{H}=t\sum_{j=1}^{L}(\hat{c}_{j}^{\dagger}\hat{c}_{j+1}+\text{h.c.})+\sum_{i\neq j}^{L}V_{ij}\hat{n}_{i}\hat{n}_{j},
\end{equation}
where $\hat{c}_j$ ($\hat{c}_j^{\dagger}$) and $\hat{n}_{j}=\hat{c}_j^{\dagger}\hat{c}_j$ denote the annihilation (creation) operator of an atom and  the particle number operator at the $j$-th site $(j=1,2,3,\cdots L)$,
$t$ the hopping coefficient,
$L$ the total number of lattice sites,
and
$V_{ij}= M_{i}M_{j}/\Delta_{c}=V_{0}\cos(i\pi\beta+\phi)\cos(j\pi\beta+\phi)$ the quasiperiodic interaction strength [see Fig.~\ref{fig:model}(c)] with $M_{i}=M_0\cos(i\pi\beta+\phi)/\sqrt{L}$, $\beta=(\sqrt{5}-1)/2$, and $\Delta_c=\omega_p-\omega_c$. Here $\phi$ is the relative phase between the background lattice and the cavity field
which can be random in experiment due to the independence of the two involved laser beams.
The randomness of $\phi$ provides us a source of statistical fluctuations in the interaction term which leads to an ensemble of spectra used in the level statistical analysis.
The connection of our model to the spin model can be found in Appendix \ref{appendix2}.

The extensive nature of the system energy is guaranteed by observing that the interaction strength $V_{ij}$ is inversely proportional to the system size, i.e., $V_{ij}\propto 1/L$
based on the fact that the single-photon Rabi frequency $M_{ij}$ is inversely proportional to the square root of the cavity size~\cite{hauke2015many}.
Such an observation shows one of the distinctive features in taking the thermodynamic limit of atomic systems confined in an optical cavity.
In reality, a cavity-confined quantum gas has a finite system size, typically in the scale of micrometers, usually corresponding to tens of lattice sites~\cite{ritsch2013cold,mivehvar2021cavity}.
In this work, we deal with a finite-sized system with $L$ ranging from 6 to 14 {\color{black} and fairly compare these numerical results under the same interaction strength $V_{0}=M_{0}^{2}/\Delta_{c}L$.}

\section{Eigenstate thermalization hypothesis and effective temperature}\label{secETH}

\begin{figure*}[tbp]
\includegraphics[width=16cm]{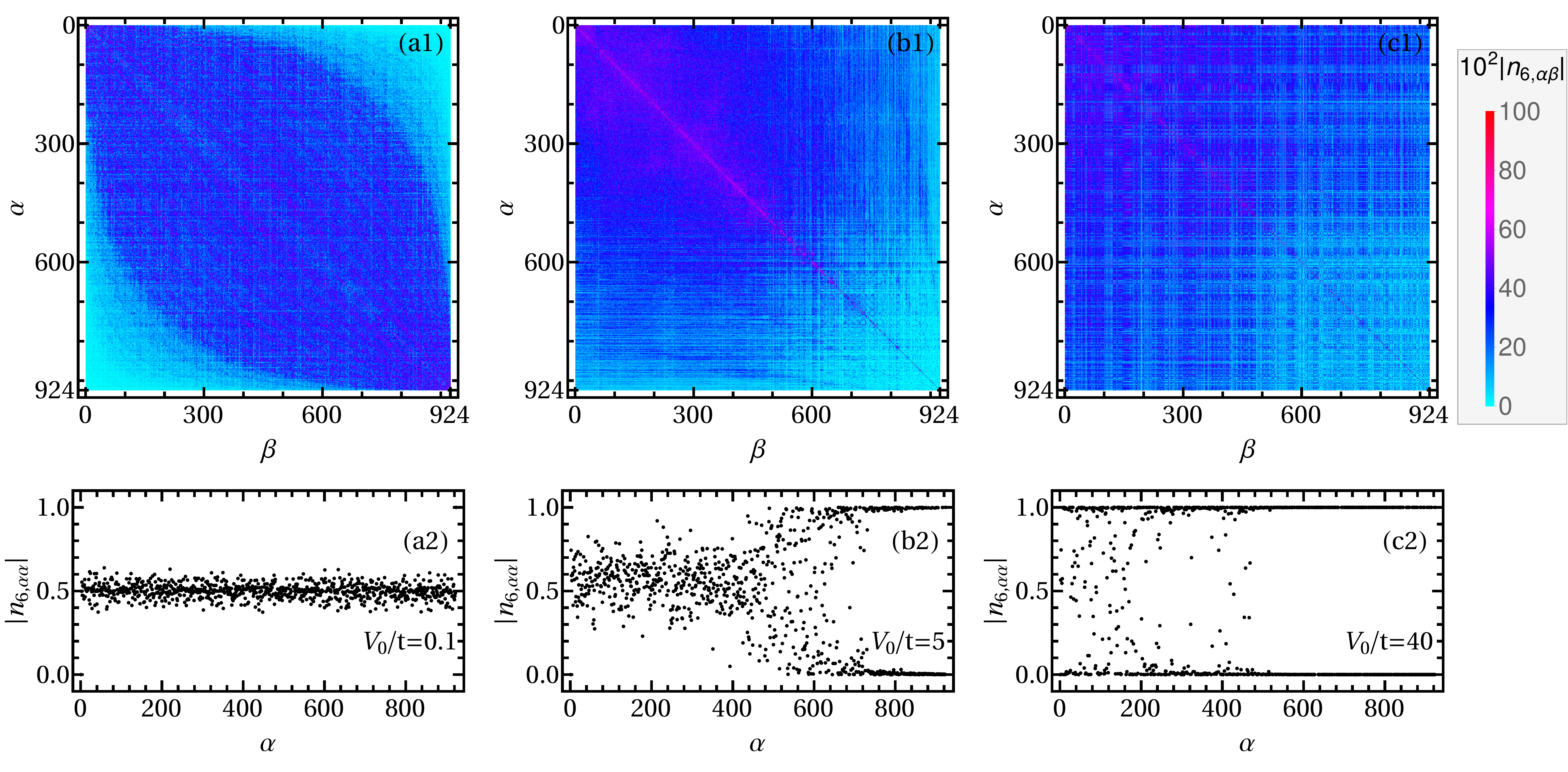}
\caption{(Color online) Distributions of the matrix element of the local particle number operator $n_{6,\alpha\beta}$ for the 6-th lattice site (central site) calculated using eigenstates (indexed by $\alpha$ and $\beta$). The lattice length is $L=12$. The interactions are $V_{0}/t=$0.1 (a1,a2), 5 (b1,b2), and 40 (c1,c2). Panels (a2), (b2), and (c2) are the diagonal elements ($\alpha=\beta$) extracted from Panels (a1), (b1), and (c1), respectively.}
\label{fig:ETH}
\end{figure*}

Ergodicity, as the essential underlying mechanism for classical thermalization, ensures that the
thermal equilibrium of a classical system is independent of the initial state.
While for an isolated quantum system, the unitary nature of the dynamics fixes the population of the evolved system over various eigenstates.
Although there exist some numerical and experimental evidences for the thermalization of isolated quantum systems after long-time dynamics~\cite{rigol2012alternatives,rigol2008thermalization},
it is still an open question whether the thermal state derived from the isolated quantum systems is consistent with that of the classical prediction in general.
The ETH~\cite{ETH1991PRA,srednicki1994chaos,nandkishore2015many,Review2016,Deutsch_2018}, which is introduced to provide us a route to approaching the final answer, states that any local observables are independent of the initial state after long-time dynamics.
The rigorous ETH \cite{rigol2012alternatives,ETH1991PRA,srednicki1994chaos,nandkishore2015many,Review2016,Deutsch_2018} requires both that the diagonal matrix elements of the relevant opera
tor calculated over various eigenstates change smoothly
and that the off-diagonal ones are {\color{black}much smaller compared to
the diagonal ones.} {\color{black} Those requirements ensure that the long-time average of any local observables, $\overline{O}=\lim_{t_{0}\rightarrow\infty}t_{0}^{-1}\int_{0}^{t_{0}}O(t)dt$, can be predicted by the microcanonical ensembles $\overline{O}_{\text{ME}}=\text{Tr}(\hat{\rho}_{\text{ME}}\hat{O})$ with the microcanonical density operator $\hat{\rho}_{\text{ME}}$. Moreover, the temporal fluctuation of $O(t)$ is negligible such that one has $O(t\gg t_{\text{relaxation}})\simeq\overline{O}\simeq\overline{O}_{\text{ME}}$ for each instant of time after relaxation, i.e., no time averaging is needed to evaluate the local observable $\overline{O}$ in the long time limit.}
While the ETH condition is satisfied for many systems~\cite{rigol2008thermalization,rigol2012alternatives}, it still lacks a universal proof for a generic many-body system.
For example, it is not obvious whether the ETH can be applied to the system with non-local quasiperiodic interactions as considered in this work.
However, we can have some insights based on our numerical results.
Recently, it is found that some long-range interacting systems such as trapped ions can hold strong ETH~\cite{ueda2021_arXiv}.
Figure~\ref{fig:ETH} shows the absolute value of the matrix elements of the particle number operator $|n_{6,\alpha\beta}|$ for the 6th lattice site, i.e., the middle one for the $L=12$ system, in the eigenstate basis.
Here, $\alpha, \beta = 1,2,3,4,\cdots$ marks the labeling of the eigenstates.
The first row of Fig.~\ref{fig:ETH} shows the full 2D map of $|n_{6,\alpha\beta}|$.
While the second row of Fig.~\ref{fig:ETH} corresponds to the diagonal parts of the first row.
{\color{black} From the first row of Fig.~\ref{fig:ETH}, we can see that the off-diagonal elements are greatly suppressed in our system, i.e., most of the off-diagonal elements are much smaller than the diagonal ones. To show this observation more clearly, we average over all the off-diagonal elements and obtain $\overline{|n_{6,\alpha\beta}|}=\sum_{\alpha\neq\beta}|n_{6,\alpha\beta}|/N_{\text{off}}=(0.69,0.52,0.022)10^{-2}$ for $V_{0}/t=(0.1,5,40)$ with $N_{\text{off}}$ the total number of the off-diagonal elements. For the diagonal elements, the averaged values are $\overline{|n_{6,\alpha\alpha}|}=0.5$ for all the cases. Thus, we get the ratios $\overline{|n_{6,\alpha\beta}|}/\overline{|n_{6,\alpha\alpha}|}=(0.014,0.010,0.00045)$ which shows that 
the second ETH condition is always satisfied for our system. 
From the second row of Fig.~\ref{fig:ETH}, we can see that the distributions of the diagonal elements $|n_{6,\alpha\alpha}|$ between adjacent eigenstates $\{|\alpha\rangle\}$ show smooth variation (Fig.~\ref{fig:ETH}(a2)), smooth-jumping mixed variation (Fig.~\ref{fig:ETH}(b2)), and jumping variation (Fig.~\ref{fig:ETH}(c2)) with increasing interaction strength. The disappearance of such a smooth variation feature indicates the breakdown of {\color{black}ETH}. In the strongly interacting regime (Fig.~\ref{fig:ETH}(c2)),
the $|n_{6,\alpha\alpha}|$ almost jumps between 0 and 1.
In such a limit, the eigenstates are close to Fock states or the standard spatial localized states.
Thus, the expectation value of the density operator does provide us an evidence for the validity of ETH in our system.  In the next section, we numerically show that the system is characterized by MBL in the ETH breakdown regime.}

\begin{figure*}[tbp]
\includegraphics[width=13cm]{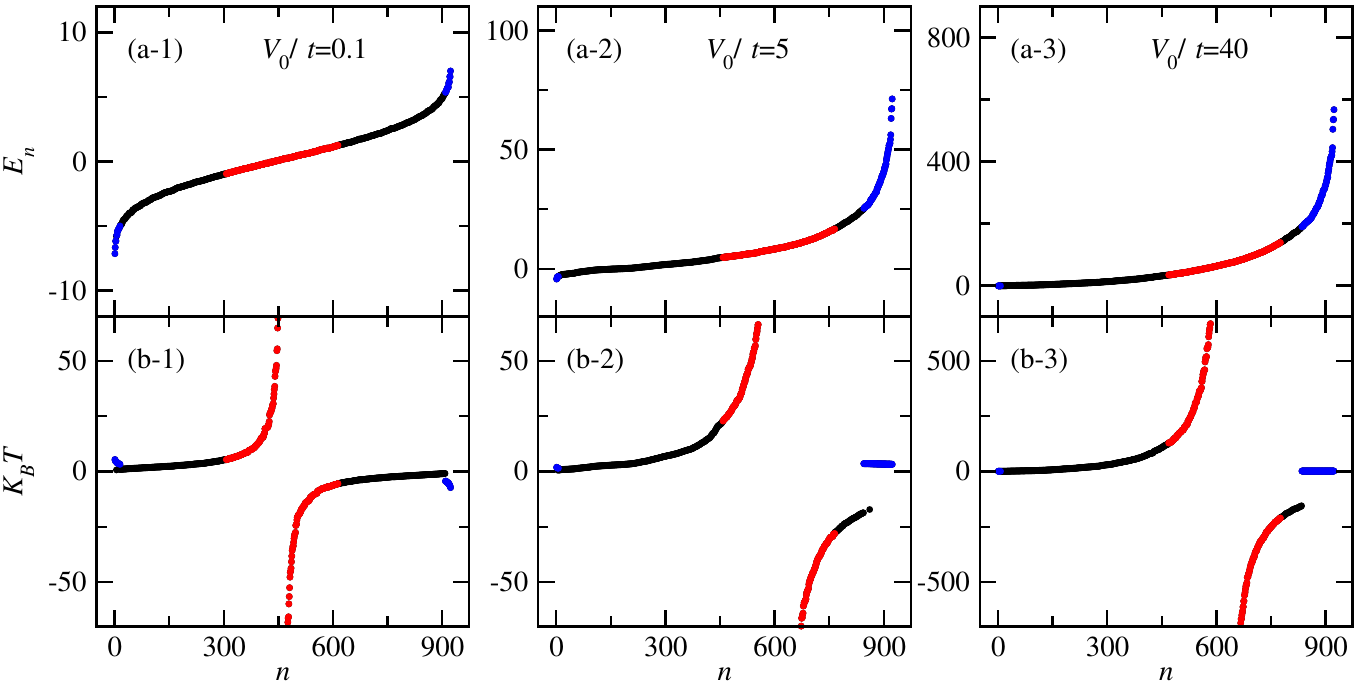}
\caption{(Color online) Effective temperatures calculated with canonical ensemble [see Eq.~(\ref{effective_temperature})]. $n$ is the index of the eigenstates sorted by eigenvalues. (a1)-(a3): the spectra for different interaction strengths. (b1)-(b3): the effective temperatures. {\color{black}The blue points are the eigenstates without well-defined effective temperatures.} The spectra marked by red color are selected in the spectral analysis in this work. The black points are eigenstates with lower effective temperatures and are dropped in the statistical analysis.
The lattice length is $L=12$.}
\label{fig:eff_temp}
\end{figure*}

\section{Numerical signatures of MBL phase}\label{secMBL}
In order to gain the insight into the localization features of the eigenstates,
we analyze the spectral statistics of the system Hamiltonian in Eq. (\ref{eq:TIH_without_a}) using the exact diagonalization method for a set of random generated $\phi$ and various system sizes $L$.   {\color{black} In the pioneering numerical MBL work \cite{oganesyan2007localization}, the existing of MBL phase was suggested in the finite temperature by numerically testing it at infinity temperature. 
In later works \cite{PRB2010MBL,yao2014many,li2016many,PRA2019Hu}, only the high effective temperature eigenstates, which usually distribute near the center of the energy spectrum, are considered in the spectral analysis. {\color{black}If the system can be described by ETH, then the effective temperature characterizes the temperature of any subsystem that is in equilibrium with the rest of the whole system \cite{nandkishore2015many}. The eigenstates with high effective temperatures are expected to correspond to 
thermal equilibrium states of high temperatures provided that ETH is applicable and thus the analysis focusing on these states is believed to provide sufficient evidences in terms of diagnosing the emergence of MBL. Mathematically}, we consider the system initialized in the $n$th eigenstate and then we can rebuild its eigenvalue $E_{n}$ by equilibrating the system to a heat bath with temperature $T_n$ which is defined via thermodynamic average \cite{li2016many} 
\begin{equation}\label{effective_temperature}
E_{n}=\frac{\text{Tr}[\hat{H}e^{-\beta_{n}\hat{H}}]}{\text{Tr}[e^{-\beta_{n}\hat{H}}]}=\frac{\sum_{m}E_{m}e^{-\beta_{n}E_{m}}}{\sum_{m}e^{-\beta_{n}E_{m}}},
\end{equation}
where we have $\beta_{n}=1/(k_{B}T_{n})$.  We call $T_{n}$ the effective temperature of the $n$th eigenstate 
.}

Figure~\ref{fig:eff_temp} shows $E_n$ (the first row) and $T_n$ (the second row) for three interaction strengths covering weak, medium and strong interaction regimes.
We can find the segment of eigenstates with effective temperature close to infinity moves toward the upper boundary of the spectra with increasing interaction strength.
It implies the portion of the spectra we selected in our analysis should not be fixed.
We choose $1/3$ of the states (marked with the red color) among the whole spectra in the following spectral analysis.

\begin{figure*}[tbp]
\includegraphics[width=16.cm]{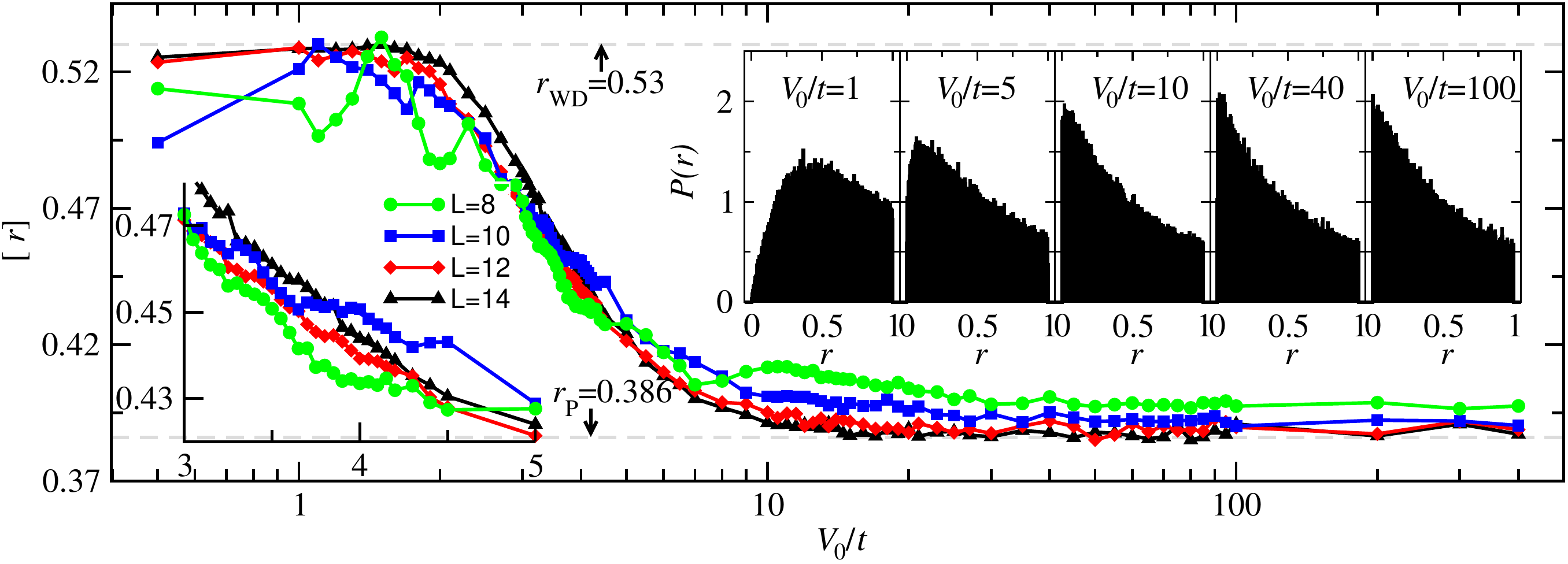}
\caption{(Color online) Spectral statistics $[r]$ as a function of interaction strength $V_0$. The $[r]$ curves for various system sizes exhibit a crossover that moves from the value of about {\color{black}$r_{\text{WD}}=0.53$} in small $V_0$ regime to that of about $r_{\text{p}}=0.386$ in the large $V_0$ regime. The two limits correspond to the {\color{black}Wigner-Dyson distribution} and Poisson distribution as illustrated in the {\color{black}upper right} insets. Here the system sizes $L=8,10,12,14$ are used. {\color{black}Lower left inset: Zoom-in plot in the regime $3\leq V_{0}/t\leq5$.}}
\label{fig:spectr}
\end{figure*}

The statistics of adjacent energy levels is one basic quantity to characterize the MBL phase transition~\cite{oganesyan2007localization}.
The dimensionless version of the adjacent energy level statistics is the adjacent min-to-max gap ratio
\begin{equation}\label{eq:rn}
r_{n}(\phi)=\frac{\text{Min}\left[\delta_{n}(\phi),\delta_{n-1}(\phi)\right] }{\text{Max}\left[\delta_{n}(\phi),\delta_{n-1}(\phi)\right]}\in[0,1],
\end{equation}
where $\delta_n(\phi)=E_n(\phi)-E_{n-1}(\phi)$ with $E_n(\phi)$ the $n$th eigenenergy in an ascending order for the system with a random phase parameter $\phi$.
Here we put the $\phi$ dependence explicitly to emphasize that all the statistical analysis on the spectrum are performed by considering the randomness of $\phi$.
We calculate the average value $[r]$ of $r_n$ via {\color{black}
\begin{eqnarray}
\label{average_rn}
[r] = \frac{1}{M}\sum_{m=0}^{M}\left[\frac{1}{N}\sum_{n=0}^N r_n(\phi_m)\right],
\end{eqnarray}}
where the $\phi_m$ are randomly generated from a uniform distribution between $0$ and $2\pi${\color{black}, i.e., one sampling is given by one phase $\phi_{m}$ which corresponds to one specific Hamiltonian $\hat{H}(\phi_{m})$.} The average value $[r]$ shown in Fig.~\ref{fig:spectr} is obtained using all of the selected levels (see the red points in Fig.~\ref{fig:eff_temp}).
\begin{figure*}[tbp]
\includegraphics[width=13cm]{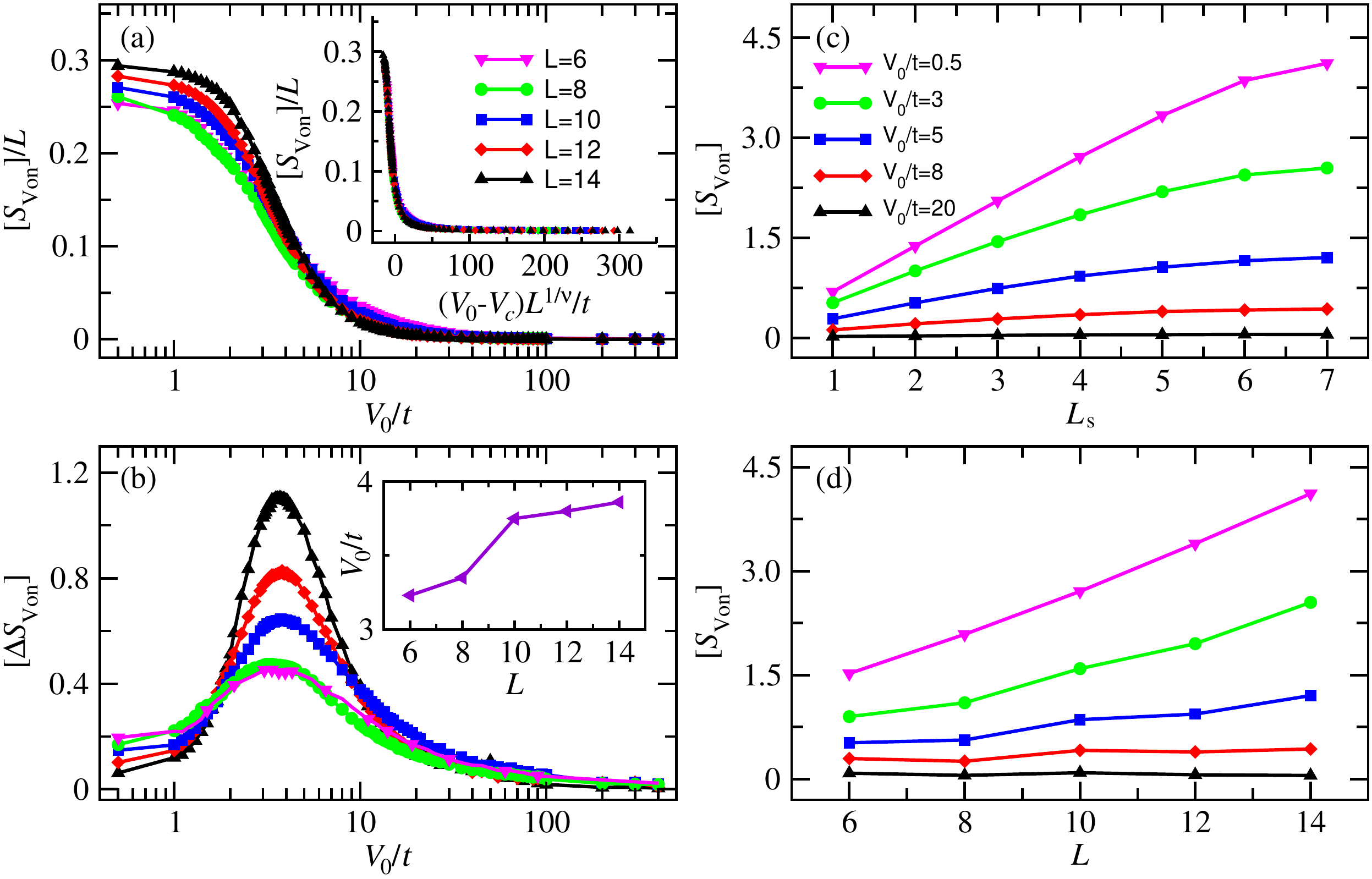}
\caption{(Color online) Entanglement entropy. (a) The average entanglement entropy per site $[S_{\text{Von}}]/L$ and (b) the statistical invariance of the entanglement entropy $[\Delta S_{\text{Von}}]$ as functions of interaction strength $V_0$.
A peak in the regime of $3<V_{0}/t<4$ can be identified in Panel (b) which corresponds to the critical point of the MBL phase.
The positions of the peaks are shown in the inset of (b).
Panel (b) shares the same legend as Panel (a).
Panels (c) and (d) show the {\color{black}$[ S_{\text{Von}}]$} as a function of the lattice length $L_s$ of the subsystem and the total system size $L$, respectively.
Panel (d) shares the same legend as Panel (c).
The definition of $[S_{\text{Von}}]$ and $[\Delta S_{\text{Von}}]$ can be found in the main text.
}
\label{fig:entangle_entrop}
\end{figure*}
The number $M$ of the $\phi$ samples is 9000, 2500, 1000, 200 for $L=(8, 10, 12, 14)$, respectively.
We find $[r]$ approaches 0.386 ($0.53$) for strong (weak) interaction strength as shown in Fig.~\ref{fig:spectr} with increasing $L$
which indicates a transition to the MBL (thermal) phase.
Similar to the case with a random local potential~\cite{oganesyan2007localization},
the crossing points of various $[r]$ curves for different $L$ move toward the larger {\color{black}$V_{0}$} and smaller $[r]$ regime {\color{black}(also see the zoom-in plot in Fig.~\ref{fig:spectr})}.
Such an observation indicates the MBL phase transition is unlikely to be extracted from the finite size scaling of the spectral statistics.
Another signature of $r_n$ statistics for the MBL (thermal) phase is that it obeys the Poisson distribution ({\color{black}Wigner-Dyson distribution}).
{\color{black}The upper right inset plots of Fig.~\ref{fig:spectr} show that the distribution of $r_n$ changes from Wigner-Dyson distribution to the Poisson distribution which
signals the transition from the thermal phase to the MBL phase with increasing $V_0$.}
\begin{figure*}[tbp]
\includegraphics[width=13cm]{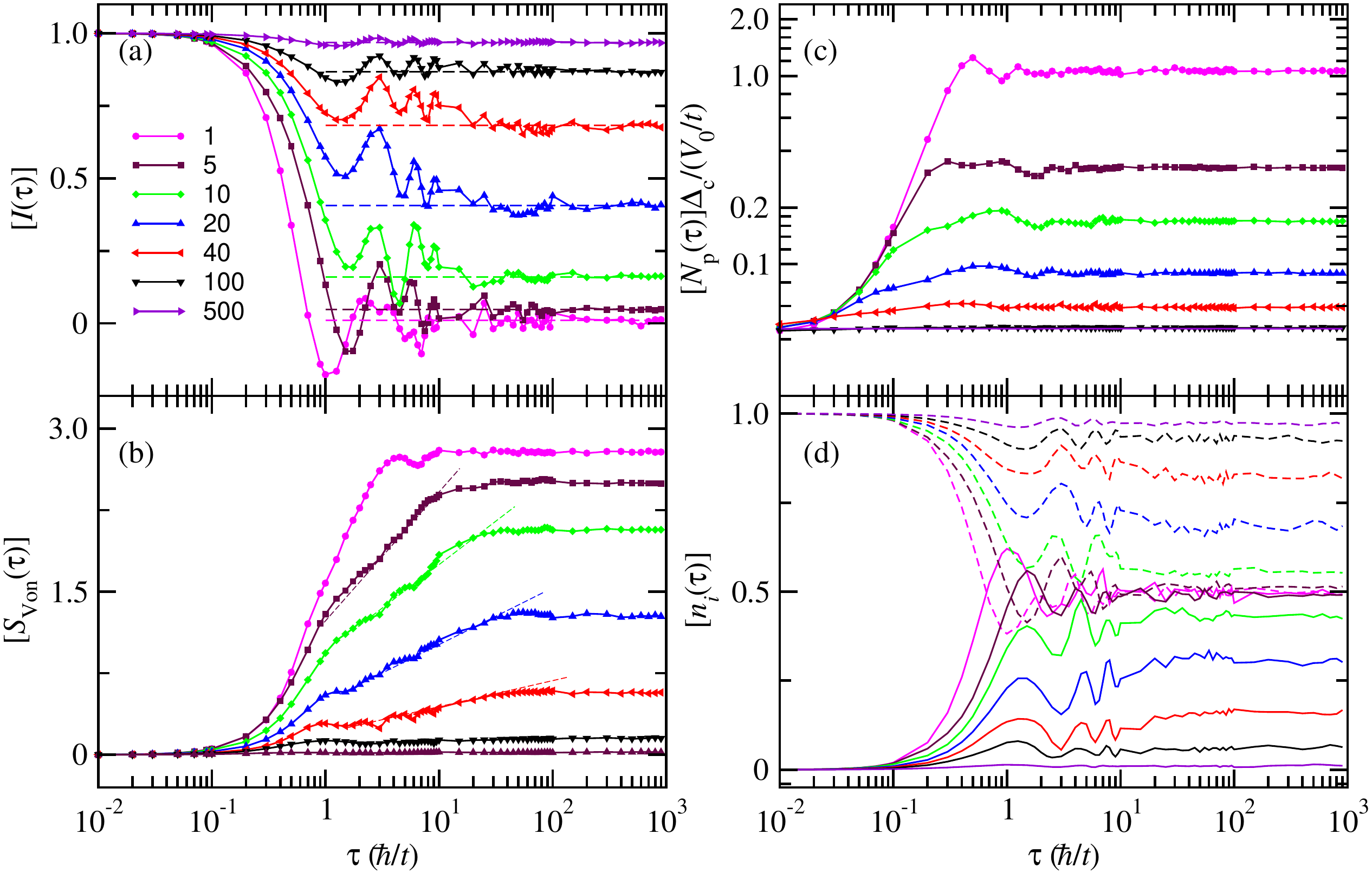}
\caption{(Color online) The time evolution of density imbalance ratio $I$ (a), photon numbers $N_{p}$ (b), entanglement entropy $S_{\text{Von}}(\tau)$ (c) and density for lattice sites $n_{5}$ (solid lines) and $n_{6}$ (dashed lines) (d) for the charge-density-wave initial state. Panels (b) and (c) share the same legend as Panel (a). $I$ saturates to a finite value in the long-time limit which is guided by the dashed horizontal lines.
The dashed lines in Panel (c) is used to indicate the logarithmic increase in $[S_{\text{Von}}]$ which typically features the MBL phases.
The system size $L=10$ is used here. }
\label{fig:time_evol}
\end{figure*}

Bipartite entanglement entropy, as another popular measure to signal the MBL phase transition, can be calculated by dividing the whole system into two subsystems $S$ and $\bar{S}$.
The Von Neumann entanglement entropy for $S$ and $\bar{S}$ reads $S_{\text{Von}}=\text{tr}(\rho_{s}\ln\rho_{s})$, where $\rho_{s}$ is the reduced density matrix of the subsystem $S$.
Since the localized states are close to Fock states (well described by the localized Wannier orbitals),
the two spatially far-separated states are weakly entangled to each other.
Therefore, the entanglement entropy obeys the area law in the MBL phase and the volume law in the thermal phase for either varying the size of the entire system or shifting the boundary between $S$ and $\bar{S}$~\cite{nandkishore2015many}.  {\color{black}Specifically, in the noninteracting limit, the bipartite entanglement entropy is close to the value $(L\ln2-1)/2$, which is the classical entropy at infinite temperature\cite{PRL1994Pages}.}
Figures~\ref{fig:entangle_entrop} (a-b) show $S_{\text{Von}}$ and its statistical variance $\Delta S_{\text{Von}}=\sqrt{\sum_{\alpha}( S_{\text{Von},\alpha}-\bar{S}_{\text{Von}})^{2}/(D-1)}$.
Here the parameters $\alpha$ and $D$ correspond to the index of the selected eigenstates and the system dimension, respectively.
The mean value $\bar{S}_{\text{Von}}$ of  $S_{\text{Von}}$ is also calculated using the same selected eigenstates.
Similarly, we also estimate the sample average $[S_{\text{Von}}]$ and $[\Delta S_{\text{Von}}]$ for the Von Neumann entanglement entropy and its fluctuation over randomly generated $\phi$.
In the weak (strong) interaction regime,
the extended (localized) eigenstates dominate the energy spectrum which yields high (low) Von-Neumann entanglement entropy per site $[S_{\text{Von}}]/L$ and hence corresponds to the thermal (MBL) phase.
In between, the numbers of extended eigenstates and localized eigenstates are comparable which leads to the peak in $[\Delta S_{\text{Von}}]$.
This observation can be used to characterize the MBL phase transition. {\color{black}In the inset plot of Fig.~\ref{fig:entangle_entrop} (a), we provide the finite size scaling analysis for the entanglement entropy. Here we take $S_{\text{Von}}(L,V_{0})=Lf[(V_{0}-V_{c})L^{1/\nu}/t]$ as scaling function and  extract the critical interaction strength $V_{c}/t=5.3\pm 1.9$ and the critical exponent $\nu=2.2\pm 1.7$ based on the collapsing behaviour shown in the inset plot of Fig.~\ref{fig:entangle_entrop} (a).}
Figure~\ref{fig:entangle_entrop} (b) and the inset indicate the critical point of our model is located in the regime of $3<V_{0}/t<4$, {\color{black} which is consistent with the estimation regime of critical interaction strength $V_{c}/t$.}
Figures~\ref{fig:entangle_entrop} (c-d) show that $[S_{\text{Von}}]$ obeys the volume law in the thermal phase (weak interaction) and the area law in the MBL phase (strong interaction) with either varying the system size $L$ [Fig. \ref{fig:entangle_entrop}(c)] or  varying the boundary between $S$ and $\bar{S}$ [Fig. \ref{fig:entangle_entrop}(d)].

\section{Time evolution}\label{timevolution}
The MBL phase manifests itself in the memory of the initial condition after a long-time dynamics~\cite{nandkishore2015many}.
Here the density imbalance ratio,
\begin{equation}
\label{eq:imbalance}
I(\tau)=\frac{N_{1}(\tau)-N_{0}(\tau)}{N_{1}(\tau)+N_{0}(\tau)},
\end{equation}
is employed to characterize the memory of the initial density distribution.
Here $N_{1}(\tau)$ [$N_{0}(\tau)$] is the number of particles at the initially occupied (unoccupied) lattice sites.
The bracket is also used here to denote the sample average over $\phi$.
As shown in Fig.~\ref{fig:time_evol}(a),
the $[I(\tau)]$ of the charge density wave (CDW) state initially oscillates in the range of several tunneling time, much shorter than the oscillation time in the case of random short-range interaction~\cite{li2017statistical}.
We see that $[I(\tau)]$ finally drops to zero in the weak interaction limit and rises to the initial value in the strong interaction limit which corresponds to the thermal phase and the MBL phase, respectively.
Another indicator of MBL is the logarithmic spreading behavior of the $[S_{\text{Von}}]$  for an initial product state~\cite{nandkishore2015many}.
As shown in Fig.~\ref{fig:time_evol}(b),
the $[S_{\text{Von}}]$ saturates after two periods especially in the large $V$ regime.
For the case with mediate strong interaction, the growth of the entanglement entropy is logarithmic before reaching saturation which is one of the signature of MBL phase.
In the strong-interaction limit,
we find the saturation value of $[S_{\text{Von}}]$ almost vanishes, namely, the entanglement entropy has no space to increase at all.
As shown in Figs.~\ref{fig:time_evol}(c) and (d),
the MBL phase also manifests itself in the conservation of the number $N_{p}$ of photons in the cavity [see Panel (c)]
and the conservation the local atomic density $n_{i}$ in the long-time evolution [see Panel (d)].
We would like to emphasize that the stability of $N_{p}$ in the MBL phase may inspire two potential applications.
It can not only be utilized to nondestructively diagnose the MBL phase by monitoring the number of leaking photons from the cavity, but leveraged for constructing a device to produce a stable number of photons~\cite{muldoon2012control}.

\section{Summary}\label{secVI}
In summary, we study the localization properties of one-dimensional degenerate Fermi gases confined in an optical cavity which effectively leads to all-to-all quasiperiodic interactions.
Using exact diagonalization,
we find that the eigenstate thermalization hypothesis is still applicable although the cavity-induced interaction is nonlocal
and that the spectrum corresponding to the effective infinite temperature is shifted for varying interaction strengths.
We also prove that the system experiences a MBL transition for increasing atom-cavity coupling strength.
The signatures of MBL phase are captured by analyzing the spectral properties, entanglement entropy, and time evolved observables.
In this work, we mainly focus on the finite-size system which is typically the case for the current experiments of intra-cavity quantum gases.
Our work may inspire the potential applications to nondestructively detect MBL phase with photons leaking from the cavity and to produce a stable number of photons with atomic gases in the MBL phase.

\begin{acknowledgments}
The authors thank Yu Chen, Wei Yi, Peng Zhang, and Hui Zhai for helpful discussions. This work is supported by the National Natural Science Foundation of China under Grant No. 12104210 (J. J.), the China Postdoctoral Science Foundation under Grant No. 2022M711496 (J. J.), the National Natural Science Foundation of China under Grant No. 11904228 and the Science Specialty Program of Sichuan University under Grand No. 2020SCUNL210 (J.-S. P.).
\end{acknowledgments}

\global\long\def\id{\mathbbm{1}}
\global\long\def\ui{\mathbbm{i}}
\global\long\def\ud{\mathrm{d}}

\setcounter{equation}{0}
\setcounter{figure}{0}
\setcounter{section}{0}
\setcounter{table}{0} 
\renewcommand{\theparagraph}{\bf}
\renewcommand{\thefigure}{S\arabic{figure}}
\renewcommand{\theequation}{S\arabic{equation}}

\onecolumngrid
\flushbottom
\appendix
\section{Derivation of the Hamiltonian with all-to-all quasiperiodic interaction in Eq. (\ref{eq:TIH_without_a})}\label{appendix1}
{\color{black}
Our strategy of realizing the 1D lattice system with all-to-all quasiperiodic interaction as described in Eq.~\eqref{eq:TIH_without_a} is schematically shown in Fig. \ref{fig:model}.
We propose to use cavity confined atoms in the presence of an external driving field $\Omega_p$ and a background lattice potential formed by the lattice beam $\Omega_b$. 
The pump laser $\Omega_{p}$ induces the cavity photon $\Omega_{c}$ via the superradiance effect 
and share the same wave vector $k$ as $\Omega_c$~\cite{baumann2010dicke, keeling2014fermionic, piazza2014umklapp, chen2014superradiance, pan2015topological}. 
The $\Omega_{p}$ and $\Omega_{c}$ couple the single-particle ground state $|g\rangle$ to the single-particle excited state $|e_1\rangle$.
Here, the all-to-all interaction between the confined atoms is yielded via absorbing and emitting cavity photons. 
The randomness in the interaction term is guaranteed by having the wave number $k_0$ of the lattice beam $\Omega_b$ incommensurate to that of the cavity mode $\Omega_c$ (i.e., the factor $\beta=k/k_0$ is an irrational number). 
The 1D geometry of the system can be experimentally realized by tuning the transverse optical trap strength stronger enough to avoid any excitation in the direction perpendicular to the cavity axis. 
In this Appendix, we show the derivation of the effective single-particle and many-body Hamiltonians.}
\subsection{Effective single-particle Hamiltonian}
{\color{black}
Typical wavelengths for the driving laser field, the cavity mode, and the background lattice discussed in this work are in the order of a few hundred of nanometers which are much larger than the typical size of the confined atoms. 
Therefore the electric dipole approximation can be used to derive the atom-light interaction. 
We denote the electric fields of the three lasers 
\begin{eqnarray}
\textbf{E}_{b} &=&E_{b}^{0}\hat{e}_{z}\cos(\omega_{b}t)\cos (k_{0}x),\label{Eb}\\
\textbf{E}_{p} &=&E_{p}^{0}\hat{e}_{y}\cos(\omega_{p}t)\cos (kz),\label{Ep}\\
\textbf{E}_{c} &=&E_{c}^{0}\hat{e}_{z}(\hat{a}^{\dag}+\hat{a})\cos(kx+\phi),\label{Ec}
\end{eqnarray}
where $\hat{a}$ ($\hat{a}^{\dagger}$), $\phi$, $E_{b,p,c}^{0}$, and $\omega_{b,p}$ are the annihilation (creation) operator of the cavity mode, the phase of the cavity mode $\Omega_{c}$ relative to $\Omega_b$, the field amplitudes, and the field frequencies, respectively. 
The  $E_{c}^{0}$ is related to the the cavity mirror area $S_{c}$, the cavity length $L_c$, and the vacuum permittivity $\epsilon_{0}$ via $E_{c}^{0}=\sqrt{\frac{\hbar \omega_c}{S_{c}L_{c}\epsilon_0}}$.

In the  Schr\"{o}dinger  picture, 
the single-particle Hamiltonian is given by
\begin{eqnarray}
\hat{H}_{\text{s}}=\hbar\omega_{c} \hat{a}^{\dag}\hat{a}+\sum_{i=1,2}\hbar\omega_{e_{i}}|e_i\rangle\langle e_{i}|+\left\{\left[\hbar\Omega_{c}(\hat{a}^{\dag}+\hat{a}) +\hbar\Omega_{p} \cos(\omega_{p}t) \right] |e_1\rangle\langle g| +\hbar\Omega_{b} \cos(\omega_{b}t) |e_2\rangle\langle g| +h.c\right\},
\end{eqnarray}
where the first, second, and the last term corresponds to the cavity mode of frequency $\omega_c$, the atomic energy levels $\hbar\omega_{e_{i}}$ (the energy of $|g\rangle$ is set to zero), and the atom-light coupling terms, respectively. 
The coupling strength or the corresponding Rabi frequencies read
\begin{eqnarray}
\Omega_{b} &=&\frac{\langle e_{2}| \hat{\textbf{d}}\cdot\hat{e}_{z}E_{b}^{0}\cos (k_{0}x)|g\rangle}{\hbar},\\
\Omega_{p} &=&\frac{\langle e_{1}| \hat{\textbf{d}}\cdot\hat{e}_{y}E_{p}^{0}\cos (kz)|g\rangle}{\hbar},\\
\Omega_{c} &=&\frac{\langle e_{1}| \hat{\textbf{d}}\cdot\hat{e}_{z}E_{c}^{0}\cos (kx+\phi)|g\rangle}{\hbar}.
\end{eqnarray}
To keep the notation simple, we use the same notation to denote the labeling of lasers and their corresponding Rabi frequencies here. 
In the spirit of the rotating wave approximation (RWA), the high frequency part of the laser fields can be rotated away leading to the Hamiltonian
\begin{eqnarray}
\hat{H}_{\text{s}}^{\text{RWA}}=\hbar\omega_{c} \hat{a}^{\dag}\hat{a}+\sum_{i=1,2}\hbar\omega_{e_{i}}|e_i\rangle\langle e_{i}|+\left[\left(\hbar\Omega_{c}\hat{a} +\frac{\hbar\Omega_{p}}{2}   e^{-\imath \omega_{p}t} \right)|e_1\rangle\langle g| +\frac{\hbar\Omega_{b}}{2} e^{-\imath \omega_{b}t} |e_2\rangle\langle g|+h.c. \right].
\end{eqnarray}
Further more, 
the time-dependence in $\hat{H}_{\text{s}}^{\text{RWA}}$ can be removed by transferring to the rotated frame 
\begin{eqnarray}
\hat{H}_{\text{Rot}}^{\text{RWA}} = \hat{U}_{\text{Rot}}^{\dagger}\hat{H}_{\text{s}}^{\text{RWA}}\hat{U}_{\text{Rot}}
=-\hbar\Delta_{c} \hat{a}^{\dag}\hat{a}-\sum_{i=1,2}\hbar\Delta_{e_{i}}|e_i\rangle\langle e_{i}|+\left[\left(\hbar\Omega_{c}\hat{a}+\frac{\hbar\Omega_{p}}{2}\right)|e_1\rangle\langle g|+\frac{\hbar\Omega_{b} }{2}|e_2\rangle\langle g| +h.c. \right],
\end{eqnarray}
where 
\begin{align}
\hat{U}_{\text{Rot}}=e^{-\imath \left(\omega_{p}\hat{a}^{\dag}\hat{a}+\omega_{p}|e_1\rangle\langle e_{1}|+\omega_{b}|e_2\rangle\langle e_{2}|\right)}
\end{align}
and $\Delta_{c}=\omega_{p}-\omega_{c},\Delta_{e_{1}}=\omega_{p}-\omega_{e_{1}},$ and $\Delta_{e_{2}}=\omega_{b}-\omega_{e_{2}}$.

In our setup, 
the atoms are strongly confined to the $x$-axis (the cavity axis) to realize the effective 1D system, we can evaluate $\Omega_p$ at $z=0$ without loss of generality.
The laser fields are far-off resonance from the atomic transitions which guarantees the validity of the adiabatic elimination approximation method~\cite{landig2016quantum} in removing the excited states $|e_{1,2}\rangle$ from the Hamiltonian $\hat{H}_{\text{Rot}}^{\text{RWA}}$.
With these considerations, 
the effective low-energy single-particle Hamiltonian for the atomic ground state after eliminating the atomic excited states is given by 
\begin{equation}\label{single_particle_Hamiltonian_2}
\hat{H}_{\text{eff,1b}}=\frac{P_{x}^{2}}{2M}+V_{x}\cos^{2}(k_{0}x)-\hbar\left[\Delta_c-\Omega_{0}\cos^{2}\left(kx+\phi\right)\right]\hat{a}^{\dagger}\hat{a}+\hbar\eta\left(\hat{a}^{\dagger}+\hat{a}\right)\cos\left(kx+\phi\right),
\end{equation}
where $P_{x}^{2}/2M$ is the kinetic energy and $M$ the atomic mass.
Here we have the effective single-photon Rabi frequency
\begin{eqnarray}
\Omega_{0}=\frac{|\langle e_{1}| \hat{d}_{z}E_{c}^{0}|g\rangle|^2}{\hbar^{2}\Delta_{e_1}},
\end{eqnarray}
the effective atom-cavity coupling coefficient
\begin{eqnarray}
\eta=\frac{\langle g| \hat{d}_{y}E_{p}^{0}|e_1\rangle\langle e_{1}| \hat{d}_{z}E_{c}^{0}|g\rangle}{2\hbar^{2}\Delta_{e_1}},
\end{eqnarray}
and the lattice potential 
\begin{eqnarray}
 V_{x}=\frac{|\langle e_{2}| \hat{d}_{z}E_{b}^{0}|g\rangle|^{2}}{4\hbar^{2}\Delta_{e_2}}.
\end{eqnarray}

This model reduces to the Aubry-Andr\'{e} model~\cite{aubry1980s} or the Harper model~\cite{harper1955single} in the limit without cavity back-action. Recently, Anderson-like localization has been predicted in the single-particle case~\cite{rojan2016localization}. In this work, we focus on the many-body regime where the impact of the atomic interaction induced by the cavity back-action is non-negligible.}

\subsection{Effective many-body Hamiltonian: all-to-all quasiperiodic interaction}

{\color{black}
The second-quantization many-body system Hamiltonian of model (\ref{single_particle_Hamiltonian_2}) is given by
\begin{equation}\label{second_quantization_SPH}
\hat{\mathcal{H}}=\int dx \hat{\Psi}^{\dagger}(x)\left[ \hat{H}_{\text{eff,1b}}+g\hat{\Psi}^{\dagger}(x)\hat{\Psi}(x)-\mu\right]\hat{\Psi}(x)
\end{equation}
where $g$ is related to the $s$-wave interaction strength $a_s$ via $g=4\pi\hbar^2a_s/M$, $\mu$ the chemical potential, and $\hat{\Psi}(x)$ $(\hat{\Psi}^{\dag}(x))$ the field annihilates (creates) operator. 
For simplicity, we assume that the lattice potential $V_x$ is much stronger than the effective single-photon Rabi frequency $\Omega_{0}$.
Under this assumption,
the tight-binding approximation to the system is applied, i.e., by expanding $\hat{\Psi}(x)$ and  $\hat{\Psi}^{\dag}(x)$ in the basis of Wannier functions $\{W(x-ja_{0})\}$ associated with the lowest Bloch band ($s$-band)
of the background Hamiltonian
 \begin{eqnarray}
\hat{H}_{\text{eff,1b}}^{0}=\frac{P_x^2}{2M}+ V_{x}\cos^{2}(k_{0}x).
\end{eqnarray}
The expansion reads
\begin{eqnarray}\label{expansion_W}
\hat{\Psi}(x)=\sum_{j=1}^{L}W(x-ja_{0})\hat{c}_{j},
\end{eqnarray}
where $a_{0}=\pi/k_{0}$ is the lattice constant, $L=L_{c}/a_{0}$ the number of the lattice sites, and $\hat{c}_{j}^{\dagger}$ ($\hat{c}_{j}$) the creation (annihilation) operator of an atom at lattice site $j$. 
Keeping the nearest neighbor tunneling of the single-particle terms, the on-site $s$-wave interaction, and the on-site energy induced by the cavity field, 
we get
\begin{equation}\label{second_quantization_SPH_1}
\hat{\mathcal{H}}=t\sum_{j=1}^{L}\left(\hat{c}_{j+1}^{\dag}\hat{c}_{j}+h.c.\right)+\frac{U_{s}}{2}\sum_{j=1}^{L}\hat{n}_{j}(\hat{n}_{j}-1)-\tilde{\Delta}_{c}\hat{a}^{\dag}\hat{a}+\sum_{j=1}^{L}M_{1,j}\hat{n}_{j}(\hat{a}^{\dag}+\hat{a})+(E_{\text{1b}}+\mu)N,
\end{equation}
where the first term is the tunneling term induced by the background Hamiltonian $\hat{H}_{\text{eff,1b}}^{0}$ with the tunneling coefficient $t$, 
the second term the on-site $s$-wave interaction with the local density operator $\hat{n}_j=\hat{c}_{j}^{\dagger}\hat{c}_{j}$, 
the third term the cavity field with the effective cavity detuning $\tilde{\Delta}_{c}=\hbar\Delta_{c}-\sum_{j}M_{2,j}\hat{n}_{j}$,
the forth term the atom-cavity coupling with effective on-site coupling coefficient $M_{1,j}$, 
and the last terms the constant terms with the total atom number $N$.
The coefficients in the Eq.~\eqref{second_quantization_SPH_1} are given by
\begin{eqnarray}
t&=&\int W^{*}(x-ja_{0})\left[\frac{P_x^2}{2M}+ V_{x}\cos^{2}(k_{0}x)\right]W(x-(j+1)a_{0})dx,\\
U_{s}&=&g\int |W(x-ja_{0})|^{4}dx,\\
M_{1,j}&=&\hbar\eta\int  |W(x-ja_{0})|^{2}\cos(kx+\phi)dx,\\
M_{2,j}&=&\hbar\Omega_{0}\int |W(x-ja_{0})|^{2}\cos^{2}(kx+\phi)dx,\\
E_{\text{1b}}&=&\int  |W(x-ja_{0})|^{2}\left[\frac{P_x^2}{2M}+V_{x}\cos^{2}(k_{0}x)\right]dx.
\end{eqnarray}
Here, $M_{1,j}$ and $M_{2,j}$ are randomness coefficients which arise from two aspects: one is the irrational value of the ratio $k/k_0$ which gives the on-site randomness ($M_{a,i}\neq M_{a,j}$ for any $i\neq j$ and $a=1,2$); another is the randomness of the phase $\phi$ and which leads to $M_{a,i}^{\text{(A)}}\neq M_{a,j}^{\text{(B)}}$ with any $\langle i,j\rangle$ combination for different theoretical samples or experimental measurements (A) and (B).

Usually, the time scale of the atomic dynamics ($\tau_a$) is much larger than the relaxation time of the cavity ($\tau_c$), i.e., $\tau_a\gg\tau_c$ or $\tilde{\Delta}_{c}\gg t$. 
Therefore, within the time interval $\tau_a$, the cavity photons relax to the steady state in a relatively short time scale which can be described by the adiabatic approximation
\begin{eqnarray}
\frac{1}{\tau_{a}}\int_{\tau-\tau_{a}/2}^{{\tau-\tau_{a}/2}}\dot{\hat{a}}(\tau')d\tau'&\approx&0,\\
\frac{1}{\tau_{a}}\int_{\tau-\tau_{a}/2}^{{\tau-\tau_{a}/2}}\hat{a}(\tau')d\tau'&\approx&\hat{a}(\tau).
\end{eqnarray}
More explicitly, the dynamics of the cavity mode $\hat{a}$ satisfies
\begin{equation}
\imath\dot{\hat{a}}=-(\tilde{\Delta}_{c}+\imath\hbar\kappa)\hat{a}+\sum_{j=1}^{L}M_{1,j}\hat{n}_{j}=0,
\end{equation}
with the decay rate $\kappa$ from which
the cavity mode dynamics is solved by \cite{rojan2016localization}
\begin{equation}\label{steady_state_a}
\hat{a}=\frac{\sum_{j=1}^{L}M_{1,j}\hat{n}_{j}}{\tilde{\Delta}_{c}+\imath\hbar\kappa}.
\end{equation}
Substituting Eq. (\ref{steady_state_a}) into Eq. (\ref{second_quantization_SPH_1}) and taking into account that the detuning $Delta_{c}$ is much larger than $\sum_{j=1}^{L}M_{2,j} n_{j}/\hbar$ and the cavity decay rate $\kappa$ (far off resonance condition as stated before), 
we find the effective Hamiltonian (up to a constant term)
\begin{equation}\label{second_quantization_SPH_3}
\hat{\mathcal{H}}=t\sum_{j=1}^{L}\left(\hat{c}_{j+1}^{\dag}\hat{c}_{j}+h.c.\right)+\frac{U_{s}}{2}\sum_{j=1}^{L}\hat{n}_{j}(\hat{n}_{j}-1)+\sum_{i,j=1}^{L}V_{ij}\hat{n}_{i}\hat{n}_{j}.
\end{equation}
where $V_{ij}=M_{1,i}M_{1,j}/\Delta_{c}$ is the strength of the all-to-all interaction. To get the approximately analytical expression for $V_{ij}$, we use the normalized harmonic oscillation function to replace the Wannier function
\begin{equation}\label{wnf}
W(x)\approx\left(\frac{\alpha}{\pi}\right)^{1/4} e^{-\alpha x^2/2},
\end{equation}
where $\alpha=2V_{x}k^{2}_{0}/(\hbar \omega_{b})$ is the normalized factor.
We find
\begin{eqnarray}
M_{1,i}=\frac{M_{0}}{\sqrt{L}}\cos\left(i\pi \beta+\phi\right)
\end{eqnarray}
and
\begin{eqnarray}
V_{ij}=V_0\cos\left(i\pi \beta+\phi\right)\cos\left(j\pi \beta+\phi\right),
\end{eqnarray}
where
\begin{eqnarray}
M_{0}= \sqrt{\frac{\hbar \omega_c}{S_{c}a_{0}\epsilon_0}}\frac{\langle g| \hat{d}_{y}E_{p}^{0}|e_1\rangle\langle e_{1}| \hat{d}_{z}|g\rangle}{2\hbar\Delta_{e_1}}{e}^{-\frac{k^2}{4\alpha}}
\end{eqnarray} 
and 
\begin{eqnarray}
V_0=\frac{M_{0}^{2}}{\Delta_{c}L}.
\end{eqnarray} 
\begin{figure}[tbp]
\includegraphics[width=8cm]{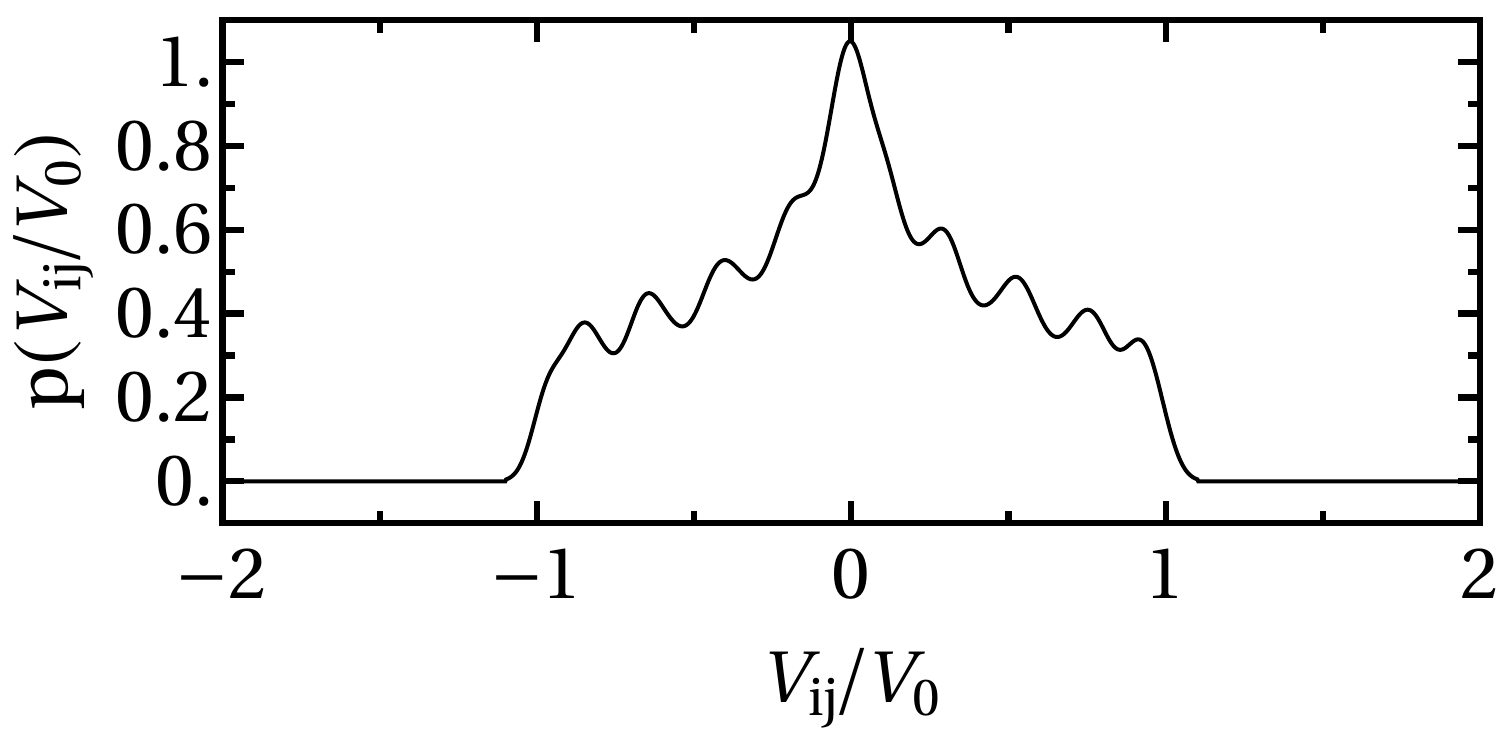}
\caption{(Color online) The probability density function of the all-to-all quasiperiodic interaction $V_{ij}/V_{0}$. Here we use 5000 samples and set lattice site number $L=14$. The relative phase $\phi$ takes an uniform distribution over $[-\pi,\pi]$ }
\label{distribution}
\end{figure}
In this work, we apply the uniform distribution to the random phase $\phi$ within the range $[-\pi,\pi]$ which gives rise to the non-uniform distribution for quasiperiodic interaction $V_{ij}/V_{0}$ (shown in Fig. \ref{distribution}). To the end, the Hamiltonian in Eq. (\ref{second_quantization_SPH_3}) can be either applied to the bosonic system or fermionic system. 
Also, the $s$-wave interaction can be tuned by the Feshbach resonance \cite{RMP2010_FR:Chin}. 
We only focus on the spinless fermionic system and turn off the $s$-wave interaction as shown in Eq. (\ref{eq:TIH_without_a}) of main text.}

\subsection{Spin model mapping}\label{appendix2}
{\color{black}
To make a connection with the literature that use spin models,
our model can be converted to a spin model using the Jordan-Wigner transformation.
Applying the Jordan-Wigner transformations $\hat{c}_{j}=\prod_{i=1}^{j-1}\left(-\hat{\sigma}_{i}^{z}\right)\hat{\sigma}_{j}^{-}$ and $\hat{c}_{j}^{\dagger}\hat{c}_{j}=\left(\hat{\sigma}_{j}^{z}+1\right)/2$ to the Hamiltonian in Eq. (\ref{eq:TIH_without_a})
and using $\hat{c}_{j}^{\dagger}\hat{c}_{j+1}=-\hat{\sigma}_{j}^{+}\prod_{i=1}^{j-1}(-\hat{\sigma}_{i}^{z})\prod_{k=1}^{j}(-\hat{\sigma}_{k}^{z})\hat{\sigma}_{j+1}^{-}=-\hat{\sigma}_{j}^{+}\hat{\sigma}_{j}^{z}\hat{\sigma}_{j+1}^{-}=\hat{\sigma}_{j}^{+}\hat{\sigma}_{j+1}^{-}$,
we arrive at a spin model (up to a constant term)
\begin{equation}
\label{eq:XXZ}
\hat{\tilde{H}}=\tilde{t}\sum_{j=1}^{L}\left(\hat{\sigma}_{j}^{+}\hat{\sigma}_{j+1}^{-}+\text{h.c.}\right)+\sum_{i\neq j}^{L}\tilde{V}_{ij}\hat{\sigma}_{i}^{z}\hat{\sigma}_{j}^{z}+\sum_{i=1}^{L}\tilde{h}_{i}\hat{\sigma}_{i}^{z},
\end{equation}
where $\hat{\sigma}_{j}^{\pm}=\left(\hat{\sigma}_{j}^{x}\pm \imath\hat{\sigma}_{j}^{y}\right)/2$ and $\hat{\sigma}_j^z$ are the standard Pauli operators, $h_{j}=\sum_{i}V_{ij}$ the effective local magnetic field, $\tilde{t}=-t/2$ the effective hopping strength, and $\tilde{V}_{ij}=V_{ij}/4$ the interaction strength, respectively.}

\twocolumngrid
\bibliography{Citations}
\end{document}